\documentclass[useAMS,usenatbib]{mn2e}
\usepackage{float}
\usepackage{graphicx}
\usepackage{grffile}
\usepackage{lineno}
\usepackage{multirow}
\usepackage{subfigure}
\usepackage{color}
\usepackage{cleveref}

\def\vFv{\nu F_{\nu}}
\def\Ep{E_{\rm p}}
\def\fermi{\it Fermi}
\def\rph{r_{\rm ph}}

\title[Are GRB Blackbodies an Artifact of Spectral Evolution?]{Are GRB
  Blackbodies an Artifact of Spectral Evolution?}\author[J. Michael
Burgess and Felix Ryde]{J. Michael Burgess$^{1,2}$\thanks{E-mail: jamesb@kth.se (JMB)} and Felix Ryde$^{1,2}$\thanks{Email: fryde@kth.se}\\
  $^{1}$The Oskar Klein Centre for Cosmoparticle Physics,
  SE-106 91 Stockholm, Sweden\\
  $^{2}$Department of Physics, KTH Royal Institute of Technology,
  AlbaNova, SE-106 91 Stockholm, Sweden}
\begin{document}

\date{Accepted XXXX December XX. Received XXXX December XX; in original form XXXX October XX}

\pagerange{\pageref{firstpage}--\pageref{lastpage}} \pubyear{2014}

\maketitle

\label{firstpage}

\begin{abstract}
  The analysis of gamma-ray burst (GRB) spectra with multi-component
  emission models has become an important part of the field. In
  particular, multi-component analysis where one component is a
  blackbody representing emission from a photosphere has enabled both
  a more detailed understanding of the energy content of the jet as
  well as the ability to examine the dynamic structure of the
  outflow. While the existence of a blackbody-like component has been
  shown to be significant and not a byproduct of background
  fluctuations, it is very possible that it can be an artifact of
  spectral evolution of a single component that is being poorly
  resolved in time. Herein, this possibility is tested by simulating a
  single component evolving in time and then folding the spectra
  through the $\fermi$ detector response to generate time-tagged event
  Gamma-ray Burst Monitor (GBM) data. We then fit both the time
  integrated and resolved generated spectral data with a
  multi-component model using standard tools. It is found that in {\it
    time-integrated} spectra, a blackbody can be falsely identified
  due to the spectral curvature introduced by the spectral
  evolution. However, in time-resolved analysis defined by time bins
  that can resolve the evolution of the spectra, the significance of
  the falsely identified blackbody is very low. Additionally, the
  evolution of the artificial blackbody parameters does not match the
  recurring behavior that has been identified in the actual
  observations. These results reinforce the existence of the blackbody
  found in {\it time-resolved} analysis of GRBs and stress the point
  that caution should be taken when using time-integrated spectral
  analysis for identifying physical properties of GRBs.
\end{abstract}


\section[]{Introduction}
Thermal emission from gamma-ray bursts (GRBs) was one of the earliest
predictions when models were first formulated to explain these extreme
astrophysical events \citep{Goodman:1986,Paczynski:1986}. In
principle, thermal emission from the jet photosphere of is a natural
explanation for the emission because the GRB jet starts its expansion
optically thick and as it evolves to lower densities the thermal
energy trapped in the flow is released. In certain scenarios, this
energy is in the form of a blackbody
\citep{Beloborodov:2010,Lundman:2014}. However, once broadband
observations of thousands of GRB prompt $\gamma$-ray spectra were
analyzed, it was found that the majority of the spectra were
non-thermal \citep{Mazets:1981,Fenimore:1982,Matz:1985ud}. There were
noted exceptions having blackbody spectra
\citep{Ghirlanda:2003,Ryde:2004te} while in others at least part of
the spectrum appeared to contain a combination of thermal and
non-thermal emission \citep{Ryde:2005bo,Ryde:2009}. Then observations
with the $\fermi$ Gamma-ray Telescope confirmed the existence of a
blackbody component in addition to the typically observed Band
component \citep{Band:1993wc} in several GRBs
\citep{Guiriec:2011jr,Axelsson:2012ic,Preece:2014ho,Burgess:2014db,Iyyani:2013dy}. The
existence of the component was shown to be statistically significant
to at least 5$\sigma$ in some cases. Additionally,
\citet{Ghirlanda:2013} showed that evidently some GRBs which have a
very narrow and hard spectrum can be well fitted by a blackbody alone
and constitute a very small fraction of all GRBs in line with earlier
observations. Even though pure blackbody spectra are rare, it is
important to note that the existence of such GRBs confirms the early
idea that emission from the photosphere does indeed play a role in
shaping GRB spectra. This strongly motivates the search for blackbody
components in non-thermal appearing GRBs as well.

The identification of this blackbody component in GRB spectra is a
crucial element in determining the origin, kinematics, structure, and
emission mechanisms of these events. For example, it has been shown
that measuring the flux of the thermal and non-thermal components
allows determination of the photospheric radius ($\rph$) and bulk jet
Lorentz factor ($\Gamma$) of the GRB jet \citep{Peer:2007}. The very
presence of the thermal component eliminates many emission models and
constrains the amount of energy that is available for the acceleration
of particles to high-energies via dissipation. It suggests the at
least part of the emission site is in the optically thick region of
the flow. Moreover, these observations are a highlight of the $\fermi$
gamma-ray space telescope mission and have enabled a greater
understanding of GRBs.

Such an important discovery warrants a deep investigation into the
reliability of the observation. As stated above, the significance of
the thermal component has been tested \citep[for
example]{Guiriec:2011jr,Axelsson:2012ic} and hence been shown not to
be a an artifact of background fluctuations. Moreover, calibration
errors have been ruled out. However, it is possible that the component
could be present in the spectra as an artifact of the spectral
evolution of a single non-thermal component. Herein, this assumption
is tested by simulating GRB pulses with known spectral evolution of a
single Band component. The pulses are fit with both Band and
Band+blackbody models in an attempt to find a blackbody where one does
not actually exist.

\section[]{General Features of the Blackbody Component}
\label{sec:bb}
While the detection of a blackbody in GRB spectra has several
implications for the physics of GRBs in itself, it is also important
to discuss how it alters the typical Band fits. In general, GRB
spectra fit with the Band function have a low-energy index ($\alpha$)
inconsistent with what is expected from the simplest synchrotron
models \citep{Preece:1998} \citep[see; however,][]{Daigne:2011}. When
a blackbody is added to the spectrum below the $\vFv$ peak, it can
alter the value of $\alpha$, possibly making it consistent with
synchrotron. This could be a resolution of the so-called synchrotron
``line-of-death'' problem. Additionally, the $\vFv$ peak ($\Ep$) of
the Band component can be shifted to higher energies, when it is fit
with a blackbody. This type of behavior has been observed in several
$\fermi$ GRBs \citep{Guiriec:2011jr,Axelsson:2012ic}. These features,
along with past theoretical predictions
\citep{Goodman:1986,Paczynski:1986,Beloborodov:2010,Lundman:2014},
strongly motivate multi-component fitting where one component is a
blackbody.

\begin{figure}
  \includegraphics[scale=1]{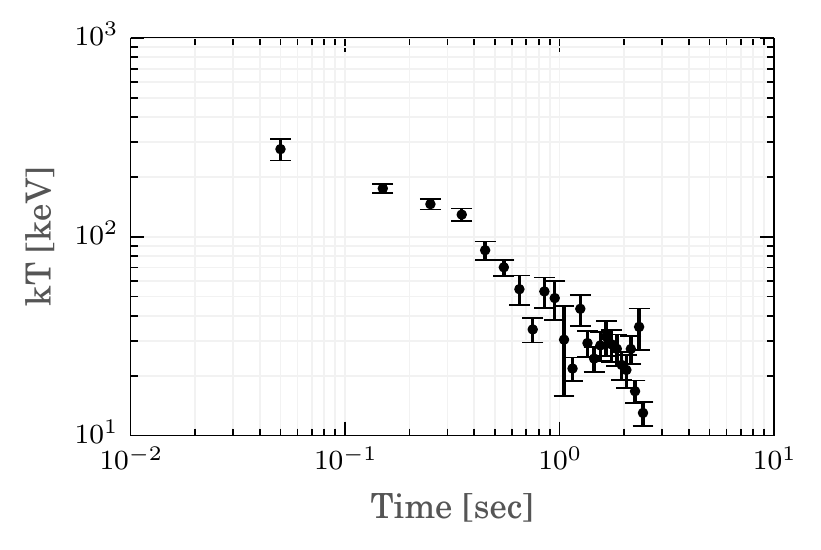}
  \caption{The evolution of $kT$ of GRB 130427A \citep{Preece:2014ho}. The
    evolution is characterized by decaying broken power law in time. }
  \label{fig:ktevoreal}
\end{figure}

The evolution of $kT$ has been well documented
\citep{Axelsson:2012ic,Ryde:2005bo,Burgess:2014db,Iyyani:2013dy} and
shown to universally decay as a broken power law in single pulses of
GRBs (see Figure \ref{fig:ktevoreal}).
Apart from the statistical significance of the blackbody component,
such a universal decay indicates a fundamental property of GRBs,
reflecting the underlying physics. For instance, it has been argued to
be a natural consequence of the relativistically expanding plasma
\citep[see, however \citet{Deng:2014ky}]{Peer:2008}. A full
understanding of the physical evolution has yet to lead to predictions
that can account for these observations.

\begin{figure}
  \includegraphics[scale=1]{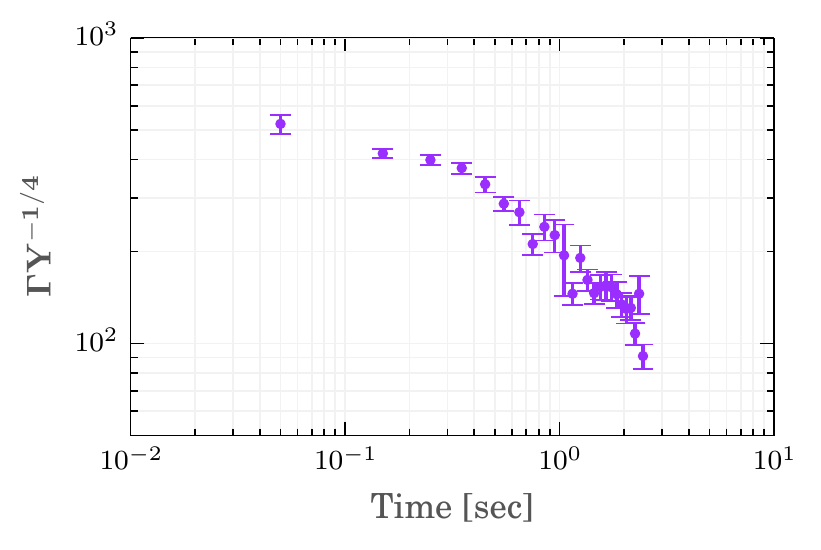}
  \caption{The evolution of $\Gamma$ in GRB 130427A
    \citep{Preece:2014ho}. The Y parameter is the ratio of total
    fireball energy to that emitted only in $\gamma$-rays.}
  \label{fig:gammaevoreal}
\end{figure}

\begin{figure}
  \includegraphics[scale=1]{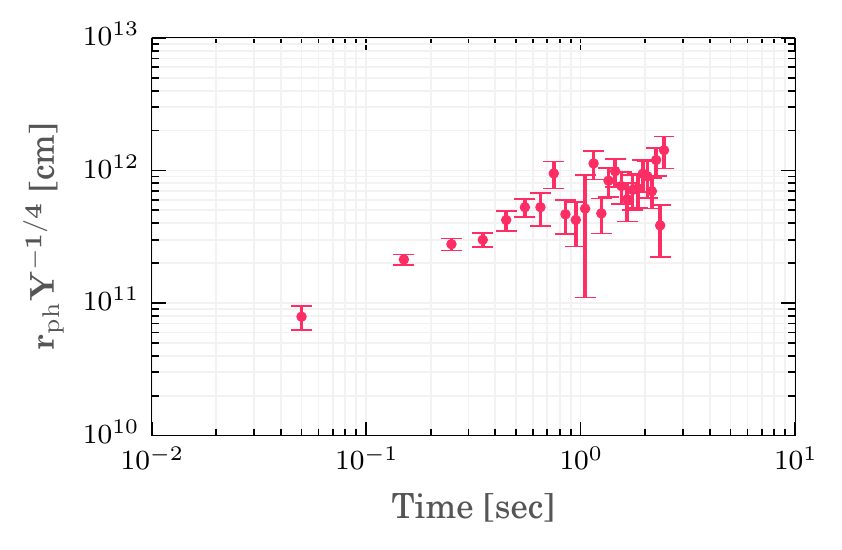}
  \caption{The evolution $r_{\rm ph}$ in GRB 130427A
    \citep{Preece:2014ho}.}
  \label{fig:rphevoreal}
\end{figure}

Perhaps the most useful and exciting feature of multi-component
fitting with a blackbody is the ability to reveal the temporal
variation of the properties of the jet such as $\rph$ and
$\Gamma$. Several studies have calculated these outflow parameters for
different GRBs \citep{Preece:2014ho,Burgess:2014db,Iyyani:2013dy,Gao:2014} and
find common trends for both parameters. As shown in
Figures \ref{fig:gammaevoreal} and \ref{fig:rphevoreal}, the photospheric radius is
seen to monotonically increase with time and $\Gamma$ is observed to
decrease with time \citep[see however][]{Gao:2014}.%
When combined with the high significance of the detection (see
Section \ref{sec:flu}) a theme begins to form that makes the detection of the
blackbody very reasonable.

There are, however; reasons to question fits with a blackbody. On
theoretical grounds, the blackbody should not appear with the simple
form%
\begin{equation}
  \label{eq:bb}
F_{\nu}\left( \varepsilon  \right)\;\propto\;\varepsilon^{3}\frac{1}{\exp\left( \frac{\varepsilon}{kT}\right)-1}
\end{equation}%
used in the above mentioned works. For instance, \citet{Peer:2011}
argue that the blackbody should appear as broadened or
``multi-colored'' by geometric and relativistic effects due to the jet
shape and high bulk Lorentz factors assumed as properties of GRBs. The
fact that pure blackbodies have been observed
\citep{Ryde:2004te,Ghirlanda:2013} places serious constraints on the
dynamics of the GRB jet as argued by
\citet{Beloborodov:2011}. For example, an unbroadened blackbody can
only be observed if the photosphere occurs during the acceleration
phase of the jet while it is photon dominated.

Moreover, the flux of the blackbody compared with the primary
non-thermal component also places severe constraints on the
dynamics. \citet{Daigne:2002dy,Zhang:2011hn} argue that introducing
magnetization into the outflow allows for some of the total energy of
the jet to be entrained in a magnetic field, reducing the available
thermal energy. Hence, the intensity of the observed blackbody would
be reduced.

These features and issues, being so constraining on the models,
require the that the existence of the blackbody be confirmed to a high
degree to advance the current understanding of GRB outflows. Hence, we
investigate the existence of the blackbody in the context of spectral
evolution of a single Band component.

\section[]{Spectral Evolution Simulations}
\label{sec:sim}
Testing for the false identification of a blackbody due to spectral
evolution requires the creation of a set of simulated GRB pulses with
known spectral evolution. The Band function, a smoothly broken power
law with low and high-energy spectral indices $\alpha$ and $\beta$
respectively ($F(E)\propto E^{\alpha,\beta}$), is the shape of the
spectrum to be simulated. This choice is justified by the fact that
the Band function is the commonly assumed spectral shape and commonly
used to fit the spectrum in both catalog work and routine analysis
\citep{Kaneko:2006,Goldstein:2012go}. A more accurate approach would
be to simulate a physical non-thermal emission spectrum from a
theoretical model evolving in a physical way in an attempt to see if a
blackbody could be reconstructed by mistake from the observations. Due
to the lack of knowledge about the emission mechanisms occurring in
GRBs, any assumption of a non-thermal model or physical model in
general would not be objective. Therefore, the Band function serves an
acceptable proxy for a single component emission spectrum.

For simplicity, the spectral evolution that will be simulated is the
classic hard-to-soft evolution typically observed in GRBs
\citep{Kargatis:1994,Band:1997}. The Band function's $\vFv$ peak, or
$\Ep$, is evolved in time as a monotonically decreasing power law with
decay index $\gamma$ (see Equation \ref{eq:ep}) while all other parameters
except the amplitude are held constant.%
\begin{equation}
  \label{eq:ep}
  \Ep\left( t \right)\;=\;E_0(t+1)^{-\gamma}.
\end{equation}
The evolution of $\Ep$ is the most pronounced change in the spectra of
observed GRBs over their respective durations. To give the simulated
pulse a GRB-like flux history, the amplitude is evolved with the KRL
pulse shape of \citet{Kocevski:2003}. With these parameters, the flux
and spectrum of the simulated pulses are completely defined as an
evolving, single component GRB pulse.

The method of simulation is described in \citet{burgess:2014b}. This
method affords the ability to map the time evolving photon spectrum
into unbinned $Fermi$ Gamma-ray Burst Monitor (GBM)
\citep{Meegan:2009} time-tagged event (TTE) data. These data are then
fitted using standard techniques. Two values of $\alpha$ ($-1$ and
$0$) are selected for each value of $\gamma=1.,1.5,2.,2.5$ in the
simulations. These values span the typically observed range of
$\alpha$ and $\gamma$. With these synthetic GRBs, a control is
established because the {\it true} spectrum in the simulations is
known. We can therefore establish the false-positive of detecting a
blackbody in a systematic way.

\section[]{The Time-Integrated Spectrum}
\label{sec:flu}
While the presence of the blackbody has been measured in the
pre-$\fermi$ era, this work will focus on the fits of Band+blackbody
made with the $\fermi$ data. The first such measurement was that of
the GRB 100724B \citep{Guiriec:2011jr} and will serve as an example
for our discussion. This GRB exhibited a blackbody in both its
time-resolved and time-integrated (fluence) spectra. The measurement
of the significance came from the time-integrated spectrum. The
spectrum was fit with both Band and Band+blackbody and the C-Stat
likelihood statistic \citep{Arnaud:2011} was used to quantify the
fit. Because the models are not properly ``nested'', the difference in
C-Stat ($\Delta_{\rm cstat}$) cannot be used to directly ascertain the
significance of adding the blackbody component
\citep{Protassov:2002gg}. Therefore simulations were made to obtain
the distribution of the $\Delta_{\rm cstat}$ statistic.

Essentially, the null hypothesis (H$_0$) is assumed to be a single
component Band function. Then several thousand simulations of the fit
of only the Band function were generated with a varying
background. These simulations were fit with both the Band and
Band+blackbody models and the distribution of $\Delta_{\rm cstat}$ was
obtained. From this distribution, it was determined that the
$\Delta_{\rm cstat}$ from the actual data was at least 5$\sigma$,
meaning that there is only one chance in 3.5 million that the
blackbody component is an \emph{artifact of background
  fluctuations}. This does not, however; test if the blackbody arises
from spectral evolution of a single component. Therefore, the
simulated GRB pulses with spectral evolution are tested here for the
presence of an artificial blackbody.

\begin{figure}
  \includegraphics[scale=1]{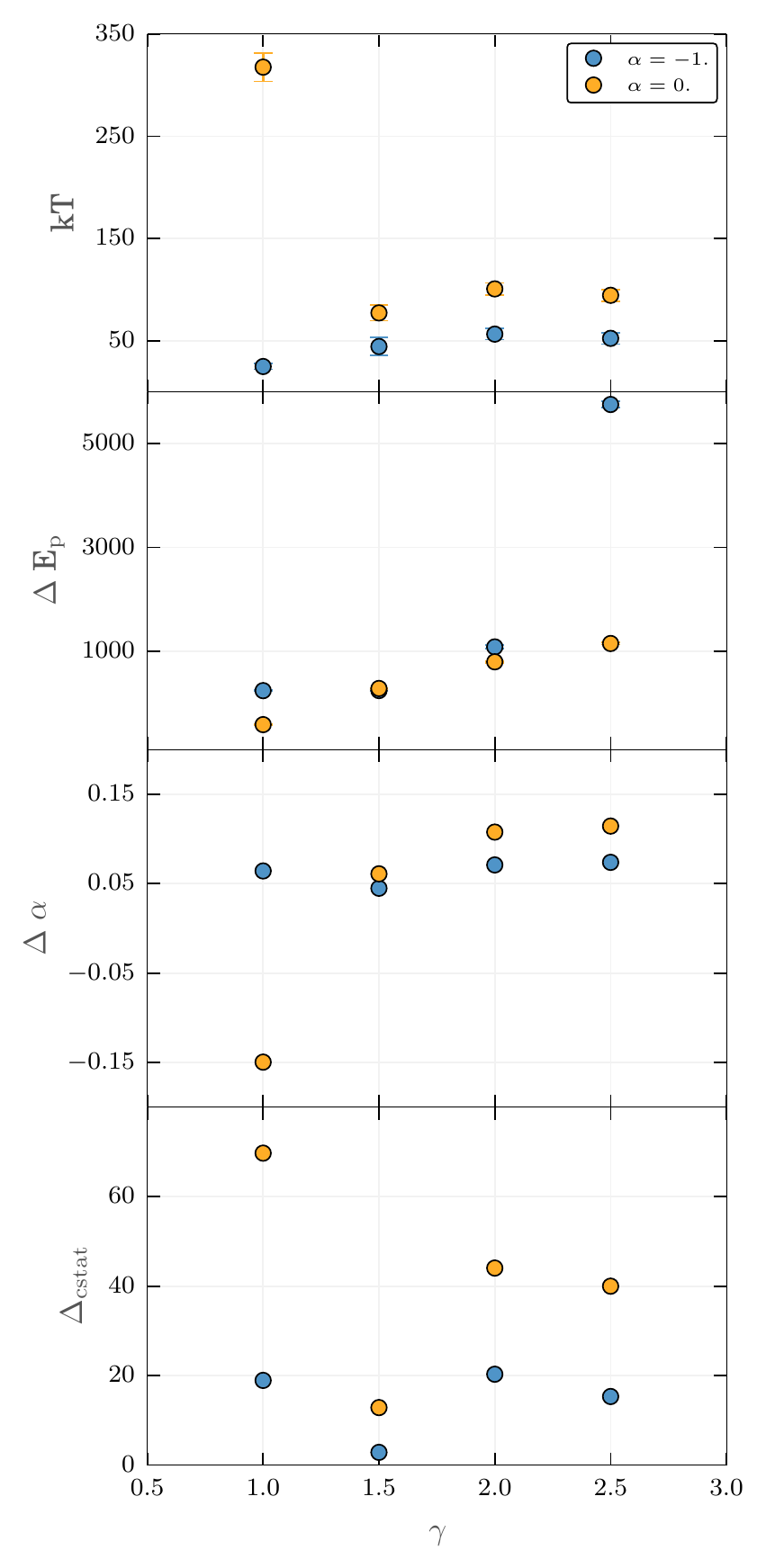}
  \caption{The properties of the integrated fits as a function of
    $\gamma$ for different values of $\alpha$. The blackbody
    characteristic energy, $kT$, appears within the range typically
    observed in the data. The shift in both $\Ep$ and $\alpha$ mimic
    the changes that have been used to justify the presence of a
    blackbody in the data. These changes should therefore not be used
    to justify the blackbody or have any physical significance
    assigned to them.}
  \label{fig:intprop}
\end{figure}
\begin{figure}
  \includegraphics[scale=1]{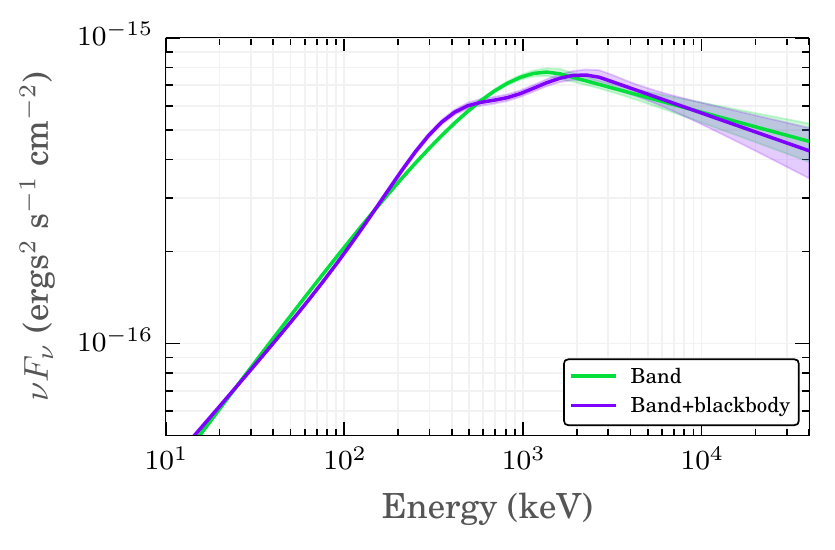}
  \caption{The $\vFv$ contour plot of a time integrated fit of the
    simulated data using the Band function and
    Band+blackbody. 1$\sigma$ contours are indicated by the shaded
    regions. The Band+blackbody model is a significantly better fit
    and captures the additional curvature below $\Ep$.}
  \label{fig:fit}
\end{figure}
\begin{figure}
\centering
  \includegraphics[scale=.5]{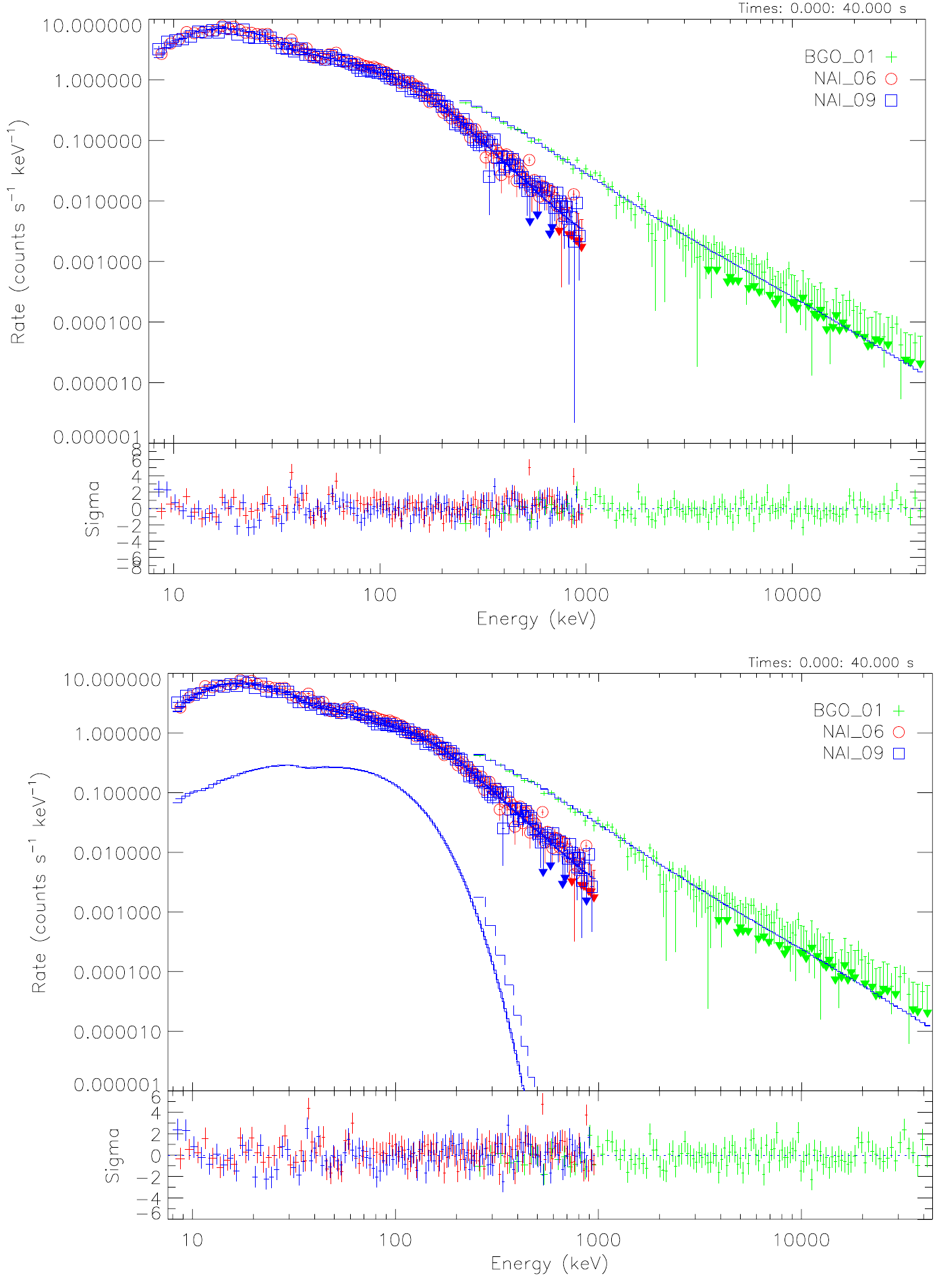}
  \caption{The convolved count spectrum of the Band function fit
    (\emph{top}) and the Band+blackbody fit (\emph{bottom}). These
    data are produced via the simulation software described in
    \citet{burgess:2014b}.}
  \label{fig:counts}
\end{figure}

Using the simulations described in Section \ref{sec:sim}, the
time-integrated spectra of the full set of parameter combinations are
fit with both the Band and Band+blackbody model (see Figures
\ref{fig:fit} and \ref{fig:counts}). The C-stat values for the fits
are recorded and their respective $\Delta_{\rm cstat}$ are calculated
yielding a tentative estimate of the significance of the addition of
the blackbody.  In all cases, a blackbody can be fit in addition to
the Band function (see Figure \ref{fig:intprop}).

\begin{figure}
  \includegraphics[scale=1]{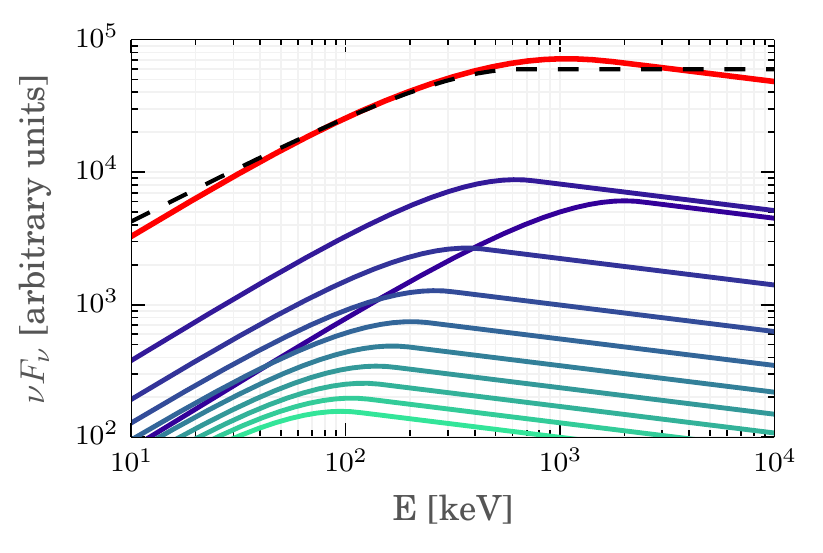}
  \caption{The simulated Band function is plotted from it's initial,
    time-resolved shape (\emph{purple}) until the end of the pulse
    (\emph{light green}). The sum of these spectra which would
    represent the integrated fit is plotted in \emph{red}. To
    demonstrate the low-energy broadening effect that spectral
    evolution introduces, a Band function with parameter values from
    the actual integrated fit is superimposed on the summed spectra
    (\emph{dashed line}). It is this excess flux at low-energy that is
    responsible for the falsely identified blackbody. It is also clear
    why including a blackbody component in the time-integrated fit
    would shift $\Ep$ to higher values.}
  \label{fig:simEvo}
\end{figure}

Moreover, the addition of the blackbody component is statistically
significant and modifies the Band function by steepening $\alpha$ and
increasing $\Ep$ in the same way that has been observed in real
data. There is a trend of increasing $kT$ as the speed of the $\Ep$
decay (increasing $\gamma$) increases. This is due to the fact that
higher values of $\Ep$ become rarer in the spectrum when $\gamma$ is
large and therefore lower $\Ep$ values in the spectra are confused as
blackbodies in the overall summed evolution (see
Figure \ref{fig:simEvo}). Faster evolution also increases the significance
of the blackbody detection.

These findings show that using fits of the time-integrated spectrum in
order to study the emission process must be done with great caution;
the spectral evolution must be accounted for. Moreover, regardless of
attempting to detect a blackbody, if the time-resolved spectra are
assumed to be Band functions, the integrated spectra can differ
significantly from a Band function leading to erroneous conclusions
about the spectral shape. In particular, justifying the significance
of any additional component, be it a blackbody or a power law, can
lead to errors which can not be addressed via statistics. This is
particularly a problem for analyzing GRBs that have highly variable
lightcurves that cannot be analyzed with time-resolved spectroscopy.

It should be pointed out that it is indeed doubtful that one should
find a pure blackbody in the integrated spectra in the first
place. Due to the evolution of $kT$ present in the time-resolved
spectra \citep{Ryde:2004te,Axelsson:2012ic}, the integrated spectra
would at best contain a multicolored blackbody due to the summing of
the evolving spectra. The fact that a blackbody is found in the
integrated spectra at all would then suggest that the blackbody
component does not evolve in time, which is difficult to consider from
a physical standpoint because the luminosity vary strongly in a
burst. Therefore, if a blackbody component does exist, the resulting
integrated spectrum should be better fit by a Band and multicolored
blackbody.

\section[]{Time-Resolved Spectra}
\label{sec:trs}

\begin{figure*}
  \centering
  \includegraphics[scale=1]{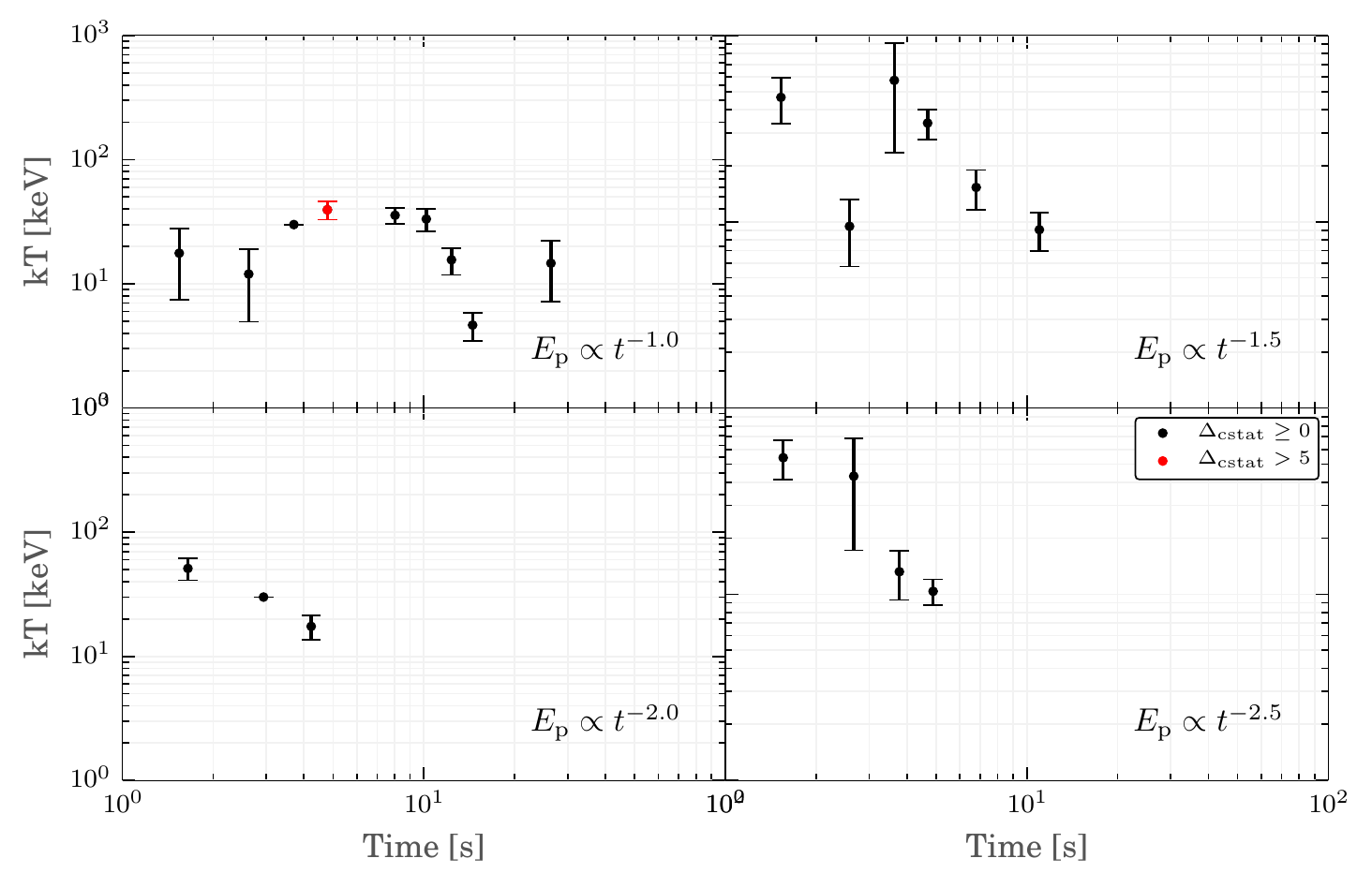}
  \caption{The evolution of $kT$ for $\alpha=-1$. Fits with a
    $\Delta_{cstat}>5$ are highlighted in \emph{red}. The \emph{blue}
    X's (see Figure \ref{fig:ktalp0}) indicate fits where the blackbody
    amplitude was negative. Therefore the only data points that would
    survive even a modest significance cut are those that are
    \emph{red} without a \emph{blue} X. Increasing the significance
    cut to reasonable values eliminates all data points.}
  \label{fig:ktalp-1}
\end{figure*}

\begin{figure*}
  \centering
  \includegraphics[scale=1]{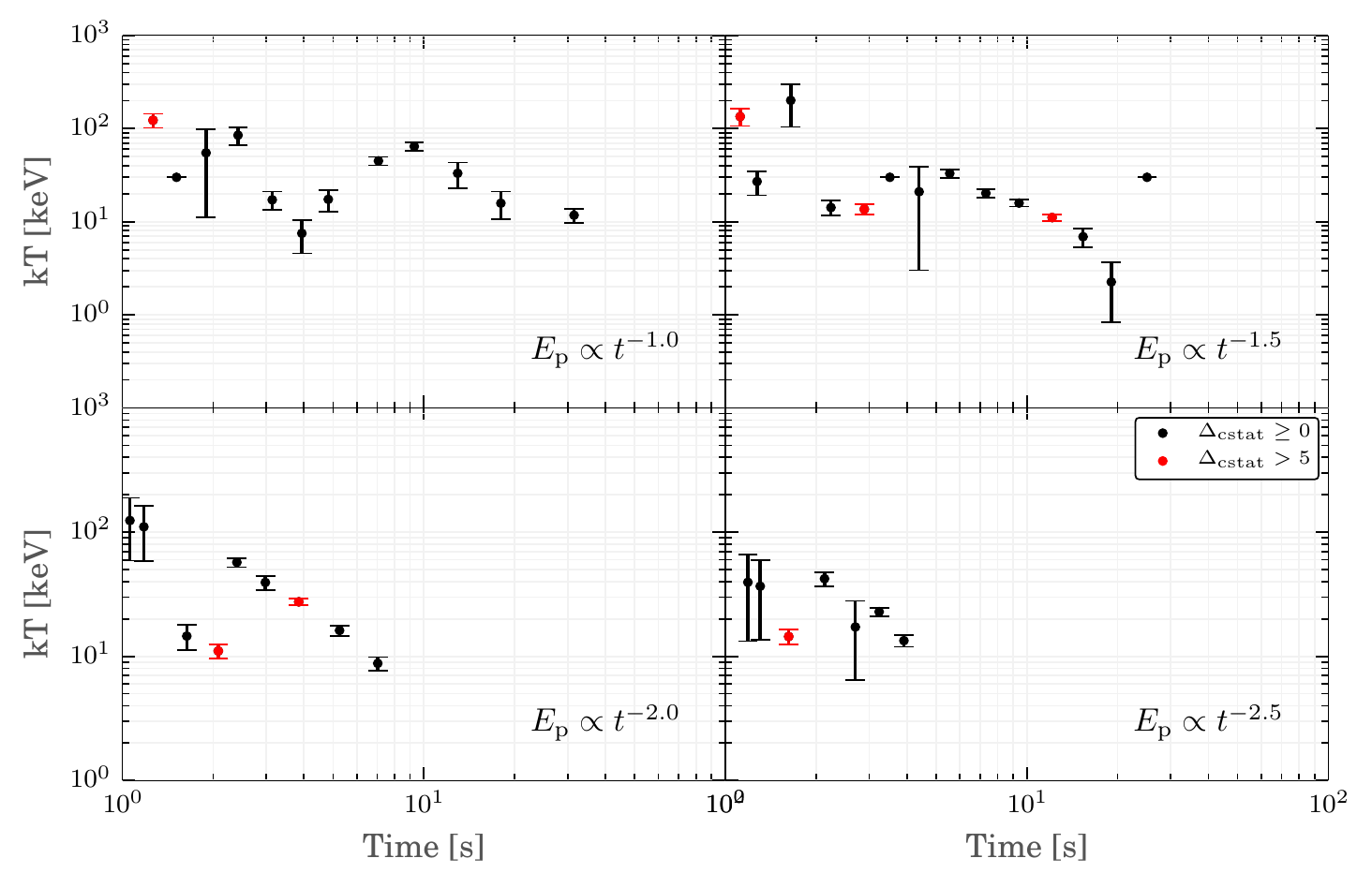}
  \caption{The same as Figure \ref{fig:ktalp-1} but for $\alpha=0$.}
  \label{fig:ktalp0}
\end{figure*}

For each simulation, the time-resolved spectra are fit with both the
Band and Band+blackbody models and the evolution of $kT$ is
computed. The fits are filtered by the improvement of the more complex
model over the simpler model via the C-Stat statistic. Figures
\ref{fig:ktalp-1} and \ref{fig:ktalp0} show the evolution of $kT$ for
various values of $\gamma$ and $\Delta_{cstat}$. There is a clear
evolution in $kT$ that is similar to the observations. This would be
troubling if the \emph{time-resolved} blackbodies found in these
simulated spectra were statistically significant. However, fitting the
more complex model to time-resolved spectra does not improve the fit
as significantly as is observed above in the time-integrated spectra.%
More importantly, many of the blackbodies found in the fits have
negative amplitudes which is not physically expected for a
photospheric component and suggests the fitting engine is simply using
the additional component to smooth over variances in the data. We
therefore assign this variation in $kT$ to the fact that $\Ep$ is
evolving and the blackbody, being a narrow spectral shape, is filling
out small, insignificant fluctuations in the spectral data below
$\Ep$. However, it is \emph{not} a required component in the spectrum
at this time-resolution.

\begin{figure}
  \includegraphics[scale=1]{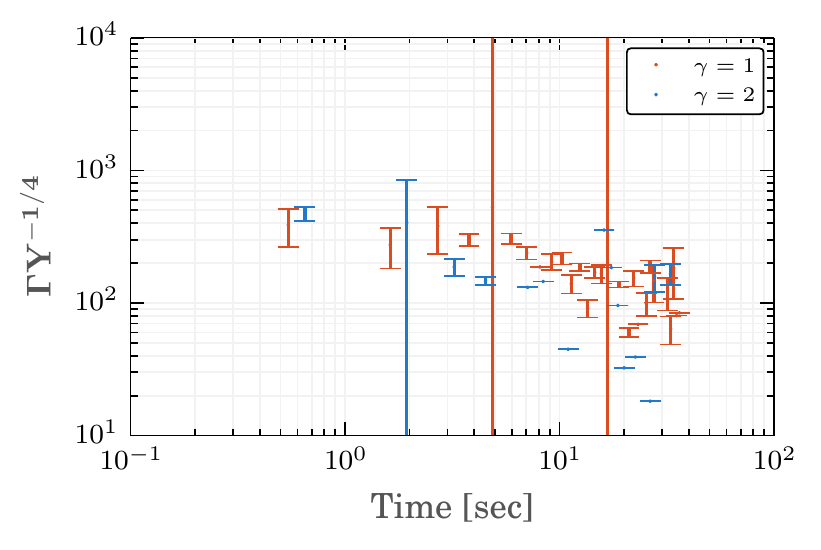}
  \caption{The inferred $\Gamma$ calculated from the falsely
    identified blackbody component. The overall behavior is not very
    different from what is observed in the data. However, no cuts on
    significance have been applied to these data points. Data points
    that would have been calculated from a blackbody with negative
    amplitude have been removed.}
  \label{fig:gamma}
\end{figure}

\begin{figure}
  \includegraphics[scale=1]{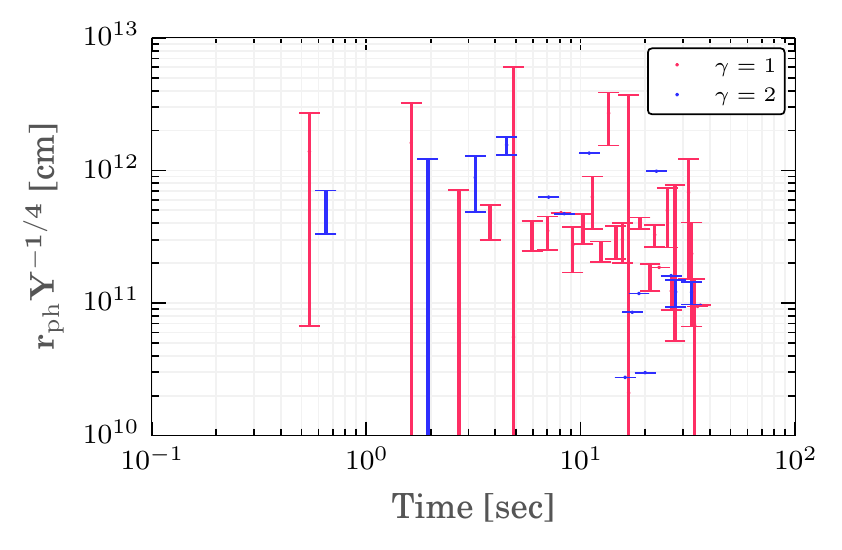}
  \caption{The inferred $r_{\rm ph}$ calculated from the falsely
    identified blackbody component. The trend is very sporadic with a
    possible decrease at late times. This is not what is observed in
    the data as shown in Figure \ref{fig:rphevoreal}.}
  \label{fig:rph}
\end{figure}

Pushing a bit further, one can use the framework derived in
\citet{Peer:2007} to compute physical jet parameters from the
multi-component fits of the simulated data. The values of $\rph$ and
$\Gamma$ are computed in Figures \ref{fig:gamma} and \ref{fig:rph}
with no significance cuts of the data points. The values of both
$\Gamma$ and $\rph$ are similar to what is typically observed in real
data, but the trends are not. The trend in $\Gamma$ does have an
overall decrease with time which is similar to what is observed in
Figure \ref{fig:gammaevoreal} but the behavior is erratic at late
times due to the value of $\Gamma$ being dependent of the total
flux. Moreover, when cuts on the significance are applied to the data
as is done in Figures \ref{fig:ktalp-1} and \ref{fig:ktalp0}, there is
no longer any trend at all because none of the blackbodies responsible
for these calculations are significant. The trend in $\rph$ is
completely dissimilar to what is typically observed. The values are
sporadic with a slight decrease at late times.

The identification of a recurring evolutionary trend for
\emph{significantly} detected blackbodies in real observations
therefore points to a physical rather than artificial presence of the
blackbody. Several authors have shown that the evolution of the
blackbody's temperature and flux have a common pattern across many
GRBs. Theoretical predictions for the evolution agree in many cases
with observations
\citep{Ryde:2005bo,Ryde:2009,Axelsson:2012ic,Iyyani:2013dy,Burgess:2014db,Preece:2014ho}. Therefore,
finding a lack of evolution of the falsely, identified blackbody in
the single-component simulations or at least one that differs from
what is observed in the real data adds support to the existence of the
blackbody in the data.

\begin{figure*}
  \centering
  \includegraphics[scale=1]{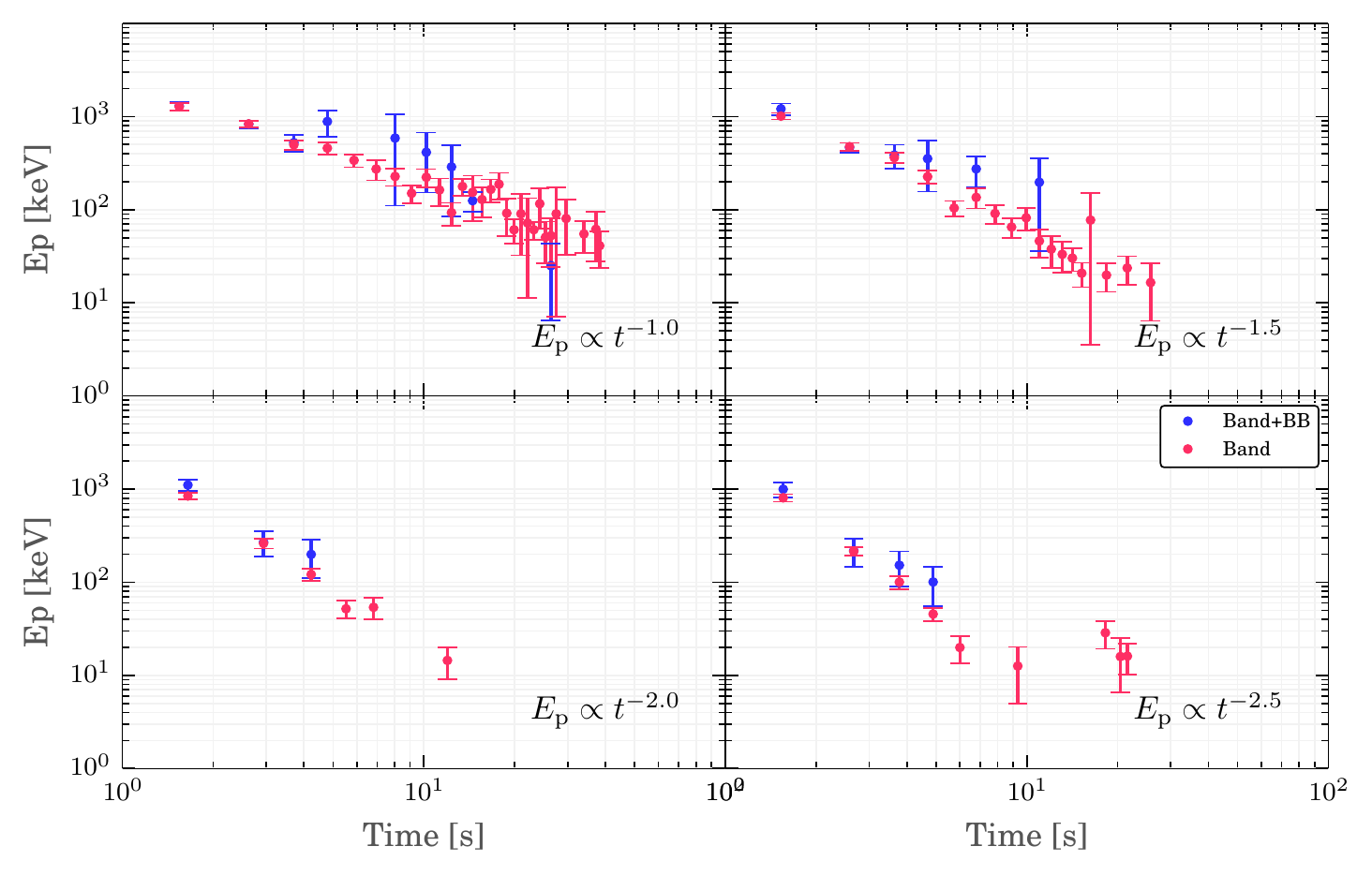}
  \caption{Band's $\Ep$ from the fits with Band and Band+blackbody for
    $\alpha=-1$. The evolution is altered near the beginning of the
    emission for some pulses when a blackbody is included in the
    model. At late times, the evolution follows the single component
    fit. Note that these fits are not filtered for significance.}
  \label{fig:epalp-1}
\end{figure*}

\begin{figure*}
  \centering
  \includegraphics[scale=1]{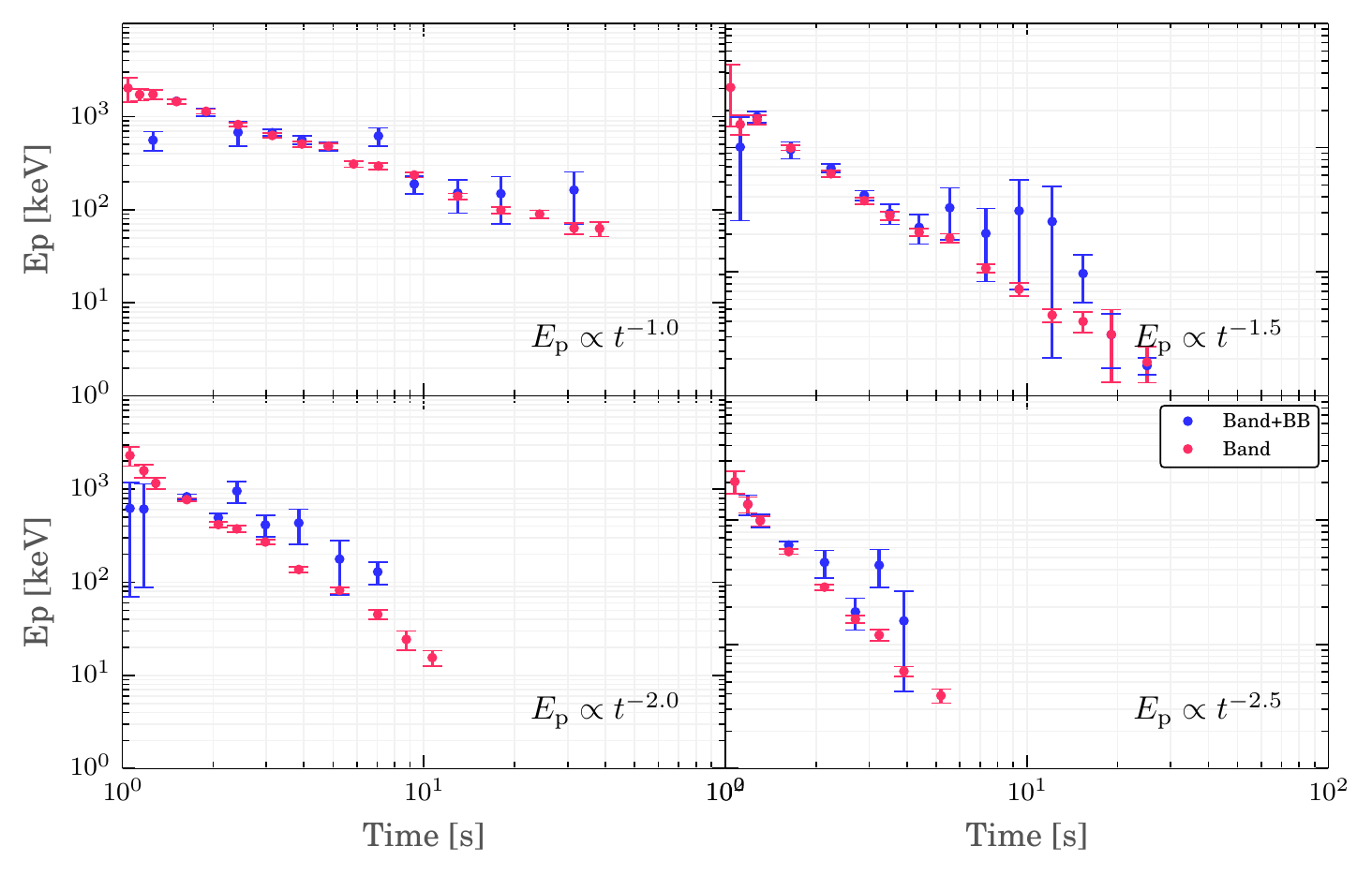}
  \caption{Same as Figure \ref{fig:epalp-1} but for $\alpha=0$.}
  \label{fig:epalp0}
\end{figure*}

One can also examine how the inclusion of the blackbody alters the
evolution of $\Ep$. As shown in \citet{burgess:2014b}, when Bayesian
blocks are used to select time intervals, the evolution of $\Ep$ can
correctly be recovered. In Figures \ref{fig:epalp-1} \ref{fig:epalp0},
the evolution of $\Ep$ from the Band and Band+blackbody fits is
compared. The evolution of $\Ep$ for the multi-component model is
altered towards the beginning of some pulses but overall it mimics the
evolutionary trend of the single component $\Ep$. The normalization is
different because $\Ep$ is shifted to slightly higher values when the
blackbody is added. This provides another clue to discerning real
blackbodies from those that are introduced by spectral evolution.


\section[]{Discussion}

The two component thermal+non-thermal fit of the GRB spectra has
become an important focus area in the field because of its power to
both constrain emission models and elucidate the structure of the GRB
jet. As discussed, it is well established that the addition of the
blackbody to spectral fits is statistically significant. The attempt
herein has been to verify that the blackbody does not appear in the
spectra as an artifact of spectral evolution of single component. We
find that in the \emph{integrated} fit an additional curvature appears
that is not accounted for by the Band function alone and it is
possible to fit a significant blackbody that is not really
there. However, in the time-resolved spectra an extra fit component
can not be significantly identified and the results do not match what
is commonly found in the actual observations. Therefore, we conclude
that the time-resolved (intra-GRB pulse) results of a significant
blackbody found previously are likely an observation of a real
additional component.

While the spectral evolution simulated in this work does not encompass
the variety of evolutions that have been observed, it does model the
most commonly observed hard-to-soft evolution. In that respect, one
can invoke Occam's razor and point to the fact that if time-resolved
analysis of a GRB with a single component Band function reveals
hard-to-soft evolution and then a multi-component analysis reveals a
significant blackbody that evolves as the expected way then it is
likely the blackbody is \emph{not} an artifact of spectral
evolution. It is entirely possible that more complex spectral
evolutions of $\Ep$ or even $\alpha$ from a single component can
reproduce the evolution of the blackbody observed in the data, but
this requires fine tuning.

It remains to be shown that physical simulations of a single component
such as synchrotron from internal shocks do not falsely produce a
blackbody in a multi-component fit. We note that indeed a single
component photosphere emission can produce complex spectra which
include a seed Planckian remnant
\citep{Peer:2005,Beloborodov:2010}. Without a clear physical model for
a single component spectral evolution, it is difficult to test the
entire parameter space that could produce an artificial blackbody from
single component evolution. Very few GRB models are advanced enough to
produce spectrally evolving lightcurves that could be folded through
the $\fermi$ instrument response and then fit with the Band+blackbody
model as is done in this paper. This test would measure the ability
for specific models that do not predict a blackbody to be mistakenly
identified as a model that does include a blackbody. For now, this
study shows that the standard hard-to-soft evolution observed in GRBs
does not produce an artificial blackbody in the time-resolved spectra.

The integrated spectra alone cannot be used to justify the presence of
a blackbody \citep[e.g. see][]{Guiriec:2011jr,Axelsson:2012ic}, but
rather the significance of the time-resolved data should be used,
which for weak and/or rapidly varying GRBs might be difficult or
impossible. As shown above, spectral evolution of a single component
Band function can result in a very significant Band+blackbody
fit. This further reinforces the point made in \citet{burgess:2014b}
that time-resolved analysis is crucial to understand the physical
properties of GRBs. The integrated properties have no direct mapping
into these physical mechanisms. The significance of the blackbody must
be evaluated in individual time bins to insure its existence. This
also highlights the importance of observed evolutionary trend in
multi-component fits. Such an observation is not as easily
quantifiable as the statistical significance of an additional
component, but it is clearly an important feature to distinguish real
(Figure \ref{fig:ktevoreal}) from artificial (Figure \ref{fig:ktalp-1}
and \ref{fig:ktalp0}) blackbodies.

The goal of this analysis has been to establish the existence of the
blackbody in light of the fact that it could arise from spectral
evolution. In the end, we have a recipe for eliminating the falsely
identified blackbodies. The following elements should be taken into
consideration:
\begin{itemize}
\item do not use the integrated spectra alone to establish the
  significance
\item the significance of the blackbody should be obtained from the
  time-resolved spectra
\item shifts in $\Ep$ and $\alpha$ do not necessarily correspond to
  evidence of the blackbody or a physical origin such as being more
  consistent with synchrotron
\item the evolution of the outflow parameters should not be erratic
  and follow the trends already demonstrated in past analysis of
  significant blackbodies found in time-resolved spectra.
\end{itemize}

\bibliographystyle{mn2e}
\bibliography{bib}

\label{lastpage}

\end{document}